\documentclass[twocolumn,superscriptaddress]{revtex4}
\usepackage{amsmath, amsfonts, amssymb, graphicx, color}

\newcommand{\<}{\langle}
\renewcommand{\>}{\rangle}
\newcommand{\beq}{\begin{equation}}
\newcommand{\eeq}{\end{equation}}
\newcommand{\bea}{\begin{eqnarray}}
\newcommand{\eea}{\end{eqnarray}}
\newcommand{\bx}{ \mathbf{x}}

\newcommand{\app}{{Appendix}}

\newcommand{\ch}{}

\newcommand{\lastequal}{Corresponding authors.}

\begin{document}

\newcommand{\deftitle}{Antigenic waves of virus-immune co-evolution}

\title{\deftitle}

\author{Jacopo Marchi}
\affiliation{Laboratoire de physique de l'\'Ecole normale sup\'erieure,
  CNRS, PSL University, Sorbonne Universit\'e, and Universit\'e de
  Paris, 75005 Paris, France}
\author{Michael L\"assig}
\affiliation{Institute for Biological Physics, University of Cologne, 50937 Cologne, Germany}
\author{Aleksandra M. Walczak}
\thanks{\lastequal}
\affiliation{Laboratoire de physique de l'\'Ecole normale sup\'erieure,
  CNRS, PSL University, Sorbonne Universit\'e, and Universit\'e de
  Paris, 75005 Paris, France}
\author{Thierry Mora}
\thanks{\lastequal}
\affiliation{Laboratoire de physique de l'\'Ecole normale sup\'erieure,
  CNRS, PSL University, Sorbonne Universit\'e, and Universit\'e de
  Paris, 75005 Paris, France}

\begin{abstract}

The evolution of many microbes and pathogens, including circulating viruses such as seasonal influenza, is driven by immune pressure from the host population. In turn, the immune systems of infected populations get updated, chasing viruses even further away. Quantitatively understanding how these dynamics result in observed patterns of rapid pathogen and immune adaptation is instrumental to epidemiological and evolutionary forecasting. Here we present a mathematical theory of co-evolution between immune systems and viruses in a finite-dimensional antigenic space, which describes the cross-reactivity of viral strains and immune systems primed by previous infections. We show the emergence of an antigenic wave that is pushed forward and canalized by cross-reactivity. We obtain analytical results for shape, speed, and angular diffusion of the wave. In particular, we show that viral-immune co-evolution generates a new emergent timescale, the persistence time of the wave's direction in antigenic space, which can be much longer than the coalescence time of the viral population.
We compare these dynamics to the observed antigenic turnover of influenza strains, 
and we discuss how the dimensionality of antigenic space impacts on the predictability of the evolutionary dynamics. Our results provide a {\ch concrete} and tractable framework to describe pathogen-host co-evolution.

\end{abstract}

\maketitle

\section{Introduction}
The evolution of viral pathogens under the selective pressure of its hosts' immunity is an example of rapid co-evolution. Viruses adapt in the usual Darwinian sense by evading immunity through antigenic mutations, while immune repertoires adapt by creating memory against previously encountered strains. Some mechanisms of in-host immune evolution, such as the affinity maturation process,  are important for the rational design of vaccines. Examples are the seasonal human influenza virus, where vaccine strain selection can be informed by predicting viral evolution in response to collective immunity\cite{Morris2018}, as well as chronic infections such as HIV \cite{Wang2015,Barton2016c,Nourmohammad2016,Nourmohammad2019}, where co-evolution occurs within each host. Because of the relatively short time scales of selection and strain turnover, these dynamics also provide a laboratory for studying evolution and its link to ecology \cite{Gandon2016}.

It is useful to think of both viral strains and immune protections as living in a common antigenic space \cite{Gandon2016}, corresponding to an idealized ``shape space'' of binding motifs between antibodies and their cognate epitopes \cite{Segel1989}. While the space of molecular recognition is high-dimensional, projections onto a low-dimensional effective shape space have provided useful descriptions of the  antigenic evolution. In the example of influenza, neutralization data from hemagglutination-inhibition assays can be projected onto a two-dimensional antigenic space \cite{Smith2004,Bedford2014,Fonville2014}. Mapping historical antigenic evolution in this space suggests a co-evolutionary dynamics pushing the virus away from its past positions, where collective immunity has developed. Importantly, the evolution of influenza involves competitive interactions of antigenically distinct clades in the viral population, generating a ``Red Queen'' dynamics of pathogen evolution \cite{VanValen1973,Yan2019}. Genomic analysis of influenza data  has revealed evolution by clonal interference \cite{Strelkowa2012a}; this mode of evolution is well-known from laboratory microbial populations \cite{Gerrish1998}. In addition, 
the viral population may split into subtypes. Such splitting or ``speciation'' events, which are marked by a decoupling of the corresponding immune interactions,  happened in the evolution of influenza B \cite{Rota1990} and of noroviruses \cite{White2014}.

The joint dynamics of viral strains and the immune systems of the host population can be modeled using agent-based simulations \cite{Ferguson2003,Bedford2012} that track individual hosts and strains. Such approaches have been used to study the effect of competition on viral genetic diversity \cite{Zinder2013}, to study geographical effects \cite{Wen2016}, and the effect of vaccination \cite{Wen2017}. Alternatively, systems of coupled differential equations known as Susceptible-Infected-Recovered (SIR) models may be adapted to incorporate evolutionary mechanisms of antigenic adaptation  \cite{Gog2002,Koelle2009,Gandon2016}.
Agent-based simulations in 2 dimensions were used to recapitulate the ballistic evolution characteristic of influenza A \cite{Bedford2012}, and to predict the occurence of splitting and extinction events \cite{Marchi2019}. In parallel, theory was developed to study the Red Queen effect \cite{Rouzine2018,Yan2019}, based on the well established theory of the traveling fitness wave \cite{Rouzine2003,Cohen2005,Desai2007b}. While effectively set in one dimension, this class of models can nonetheless predict extinction and splitting events assuming an infinite antigenic genome \cite{Yan2019}.

In this work, we propose a  co-evolutionary theory in an antigenic interaction space of arbitrary dimension $d$, which is described by joint non-linear stochastic differential equations coupling the population densities of viruses and of protected hosts. We show that these equations admit a $d$-dimensional antigenic wave solution, and we study its motion, shape, and stability, using simulations and analytical approximations. Based on these results, we discuss how canalization and predictability of antigenic evolution depend on the dimensionality $d$.

\section{Results}
\subsection{Coarse-grained model of viral-immune co-evolution}
\begin{figure}
\begin{center}
\includegraphics[width=\linewidth]{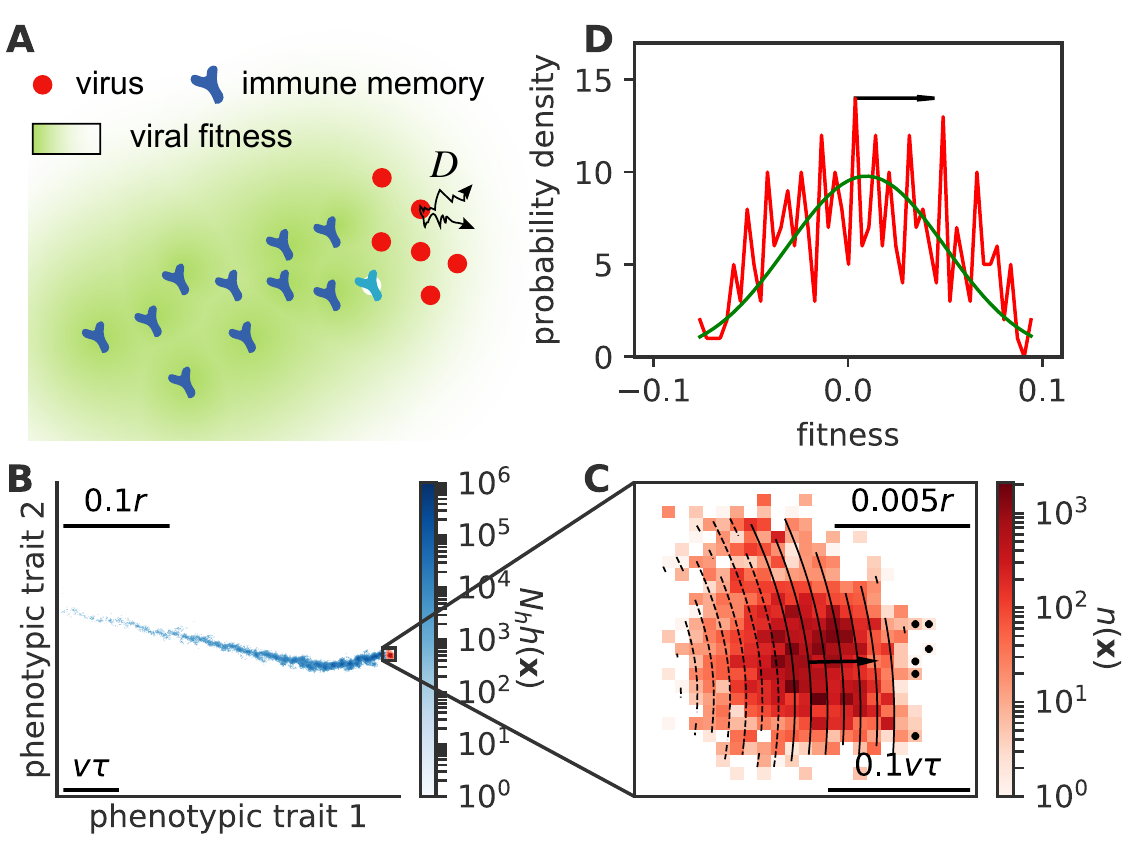}
\caption{{\bf A simple model of viral-host co-evolution predicts the emergence of an antigenic wave.}
  {\bf A.} Schematic of the co-evolution model. Viruses proliferate while effectively diffusing in antigenic space (here in 2 dimensions) through mutations, with coefficient $D$. Past virus positions are replaced by immune protections (light blue). Immune protections create a fitness gradient for the viruses (green gradient) favoring strains at the front. Both populations of viruses and immune populations are coarse-grained into densities in antigenic space.
  {\bf B.} Snapshot of a numerical simulation of Eq.~\ref{dndt}-\ref{dhdt} showing the existence of a wave solution. The blue colormap represents the density of immune protections $h(\bx,t)$ left behind by past viral strains. The current virus density $n(x)$ is shown in red.
  {\bf C.} Close-up onto the viral population, showing fitness isolines. The wave moves in the direction of the fitness gradient (arrow) through the enhanced growth of stains at the edge of the wave (black dots).   {\bf D.} Distribution of fitness across the viral population (corresponding to the projection of B. along the fitness gradient). Parameters for B-D: $D/r^2=3\cdot 10^{-9}$, $N_h=10^8$, $\ln R_0=3$, $M=1$.}
\label{fig1}  
\end{center}
\end{figure}

Our model describes the joint temporal evolution of populations of viruses and immune protections in some effective antigenic space of dimension $d$.
Both viral strains and immune protections are labeled by their position $\bx=(x_1,\ldots,x_d)$ (or ``phenotype'') in that common antigenic space (Fig.~\ref{fig1}A). In that space, viruses randomly move as a result of antigenic mutations and proliferate through infections of new hosts. Immune memories are added at the past positions of viruses. Immune memories distributed across the host population provide protection that reduces the effective fitness of the virus.
We coarse-grain that description by summarizing the viral population by a density $n(\bx,t)$ of hosts infected by a particular viral strain $\bx$, and immunity by a density $h(\bx,t)$ of immune memories specific to strain $\bx$ in the host population.

At each infection cycle, each host may infect $R_0$ unprotected hosts, where $R_0$ is called the basic reproduction number. However, a randomly picked host  is susceptible to strain $\bx$ with probability $(1-c(\bx,t))^M$, where $c(\bx,t)$ is the coverage of strain $\bx$ by immune memories of the population, and the number $M$ of immune memories carried by each host.
Because of cross-reactivity, which allows immune memories to confer protection against closeby strains, immune coverage is given as a function of the density of immune memories:
\beq\label{fit_def}
c(\bx,t)=\frac{1}{M}\int d\bx' h(\bx',t)H(\bx-\bx'),
\eeq
where $H(\bx-\bx')=\exp(-|\bx-\bx'|/r)$ is a cross-reactivity kernel describing how well memory $\bx'$ protects against strain $\bx$, and $r$ is the range of the coverage provided by cross-reactivity. In summary, the effective growth rate, or ``fitness'', of the virus is given by $f(\bx,t)\equiv {\ln[R_0(1-c(\bx,t))^M]}$.

The coupled dynamics of viruses and immune memories is then described by the stochastic differential equations  {\ch (with time in units of infection cycles throughout)}:
\begin{align}
{\partial_t n(\bx,t)}&=f(\bx,t)n(\bx,t)+D\partial_\bx^2 n+\sqrt{n(\bx,t)}\eta(\bx,t)\label{dndt}\\
{\partial_t h(\bx,t)}&=\frac{1}{N_h} \left[n(\bx,t)- N(t)\frac{h(\bx,t)}{M}\right] \label{dhdt}. 
\end{align}
Here $\eta$ is a Gaussian white noise in time and space,  $\<\eta(\bx,t)\eta(\bx',t')\>=\delta(\bx-\bx')\delta(t-t')$, accounting for demographic noise \cite{Hallatschek2011c}. This stochastic term is crucial, as it will drive the evolution of the wave.
The diffusion constant $D$ describes the effect of infinitesimal mutations on the phenotype, $D=\mu\<\delta x_1^2\>/2$, where $\mu$ is the mean number of mutations per cycle, and $\<\delta x_1^2\>$ the mean squared effect of each mutation along each antigenic dimension (assuming that mutations do not have a systematic bias, $\<\delta x_1\>=0$). The continuous-diffusion assumption implied by Eq.~\ref{dndt} is only valid when there are many small mutation effects, $\mu\gg 1$ and $\delta \bx \ll r$, {\ch in constrast with regimes where mutations are rare but have a substantial fitness effect drawn from a distribution \cite{Good2012,Rouzine2018}. Our choice is simpler in that it describes the mutation process through a single parameter $D$. Along with the choice of the cross-reactivity kernel $H$, it also naturally preserves the isotropy of the antigenic space.}

The total viral population size, or number of infected hosts, $N(t)=\int d\bx\, n(\bx,t)$ is subject to fluctuations. At the same time, the host population size $N_h$,  remains constant because newly added memories (first term of right-hand side of Eq.~\ref{dhdt}) overwrite existing ones picked uniformly at random (second term of r.h.s. of Eq.~\ref{dhdt}). Since each host carries $M$ immune receptors, we have $\int d\bx h(\bx,t)=M$.

If we assume that the system reaches an evolutionary steady state, with stable viral population size $N(t)=N$, then Eq.~\ref{dhdt} can be integrated explicitly:
\beq\label{eq:hvsn}
h(\bx,t)=\frac{M}{N}\int_{-\infty}^t \frac{dt'}{\tau}\, e^{-\frac{t-t'}{\tau}}n(\bx,t'),
\eeq
with $\tau=M N_h/N$. This equation shows how the density of protections reflects the past evolution of the viral population.

\subsection{Antigenic waves}
We simulated \eqref{dndt}-\eqref{dhdt} on a square lattice (Methods) and found a stable wave solution (Fig.~\ref{fig1}B-D). The wave has a stable population size $N$, and moves approximately ballistically through antigenic space, pushed from behind by the immune memories left in the trail of past viral strains (Fig.~\ref{fig1}B). These memories exert an immune pressure on the viruses,  forming a fitness gradient across the width of the wave  (Fig.~\ref{fig1}C), favoring the few strains that are furthest from immune memories, at the edge of the wave.

We assume that the solution of the coupled evolution equations \eqref{dndt}-\eqref{dhdt} takes the form of a moving quasispecies in a $d$-dimensional antigenic space, 
\beq
n(\bx,t)=\frac{N}{\sqrt{2\pi \sigma^2}}\exp\left[-\frac{(x_1-vt)^2}{\sigma^2}\right]\rho(x_2,\ldots,x_d). 
\label{nxt}
\eeq
Here, we have written the solution in a co-moving frame, in which a motion with constant speed $v$ takes place in the direction of the coordinate $x_1$, and fluctuations in the other dimensions, $\rho(x_2,\ldots,x_d,t)$, centered around $x_i=0$ for $i>1$, are assumed to be independent. In the next sections, we will analyse solutions of this form. First, we will project the $d$-dimensional antigenic wave onto the one-dimensional fitness space; this projection produces a travelling fitness wave \cite{Tsimring1996,Rouzine2003, Cohen2005,Desai2007b,Neher2013} that determines the antigenic speed $v$ and the mean pair coalescence time $\<T_2\>$ of the viral genealogy. Second, we will study the shape of the $d$-dimensional quasispecies and determine the fluctuations in the transverse directions. These fluctuations produce a key result of this paper: immune interactions canalize the evolution of the antigenic wave; this constraint can be quantified by characteristic time scales governing the transverse antigenic fluctuations. Canalization is most pronounced in spaces of low dimensionality $d$ and, as we discuss below, affects the predictability of antigenic evolution.

\subsection{Speed of antigenic evolution} 
Projected onto the fitness axis $f=f(\bx,t)$, the solution is approximately Gaussian (Fig.~\ref{fig1}D). This representation suggests a strong similarity to the fitness wave solution found in models of rapidly adapting populations with an infinite reservoir of beneficial mutations \cite{Tsimring1996,Rouzine2003, Cohen2005,Desai2007b,Neher2013}. To make the analogy rigorous, we must assume that the fitness gradient in antigenic space is approximately constant, meaning that fitness isolines are straight and equidistant. Mutations along the gradient direction have a fitness effect that is linear in the displacement, while mutations along perpendicular directions are neutral and can be treated independently. Note that while we will use this projection onto fitness to compute the speed of the antigenic wave, the underlying antigenic wave remains in $d$ dimensions; we will come back to transverse fluctuations in the next sections.

{\ch There are several models of fitness waves that differ in the assumptions on the statistics of mutational effects.}
Our assumption of diffusive motion makes our projected dynamics equivalent to that studied in ref.~\cite{Neher2013}, which itself builds on earlier work
\cite{Cohen2005}. {\ch This equivalence results from the two key assumptions of the mutation model in antigenic space: mutations have a small effect, and their distribution is isotropic, meaning that there are as many deleterious as beneficial mutations.}
In the limit where the wave is small compared to the adaptation time scale, $v\tau \gg \sigma$, the wave may be replaced by a Dirac delta function at $\bx=(vt,0,\ldots,0)$ in Eq.~\ref{eq:hvsn}. One can then calculate explicitly  the immune density (upstream of the wave) and coverage (downstream of the wave, using Eq.~\ref{fit_def}):
\begin{align}
\label{happrox}
 h(\bx,t)&\approx \frac{M}{v\tau}e^{-\frac{vt-x_1}{v\tau}}\Theta(vt-x_1)\delta(x_2)\cdots\delta(x_d),\\
\label{capprox}
  c(\bx,t)&\approx \frac{e^{-(x_1-vt)/r}}{1+v\tau/r},\quad x_1\geq vt,\ x_{i>1}\ll r
\end{align}
where $\Theta(x)=1$ for $x\geq 0$ and $0$ otherwise. This idealized exponential trail of immune protections $h(\bx,t)$ corresponds to the blue trace of Fig.~\ref{fig1}B, and the coverage or fitness gradient to the isolines of Fig.~\ref{fig1}C.

In the moving frame of the wave, $(u,x_2,\ldots,x_d)$, with $u=x_1-vt$, the local immune protection and viral fitness can be expanded locally for $u,x_i\ll v\tau$ (see \cite{Rouzine2018} for a similar treatment in a one-dimensional antigenic space):
\beq\label{eq:gradient}
f((u,x_{i>1});t)\approx \ln\left[ R_0 {\left(1-\frac{e^{-u/r}}{1+v\tau/r}\right)}^M\right]\approx f_0+su,
\eeq
where $f_0=\ln R_0 -M\ln[1+r/(v\tau)]$ is the average population fitness, and
\beq\label{eq:s}
s=|\partial_{x_1}f|=\frac{M}{r}\left(R_0^{1/M}-1\right)
\eeq
is the fitness gradient. Rescaling the antigenic variable $x_1$ as $sx_1$, this process is equivalent to the evolution of a population where mutation effects are described by diffusion in fitness space with coefficient $Ds^2$. This is precisely the model from which the fitness wave solution of Ref.~\cite{Cohen2005,Neher2013} was described (see \app). In the following we will use results from these works to describe the antigenic wave. {However, we note that in the usual fitness wave theory, population is kept constant by construction, which implies} that fitness is only relevant when compared to the mean of the population. By contrast, in our model population size is {itself a dynamical variable}, and fitness is defined as an absolute growth rate. {In this version of the model,} the fitness of the whole viral population undergoes continuous negative drift due to the constant adaptation of immune systems, encoded in the $-svt$ term in Eq.~\ref{eq:gradient}. This negative fitness drift has an analogous effect to subtracting the mean fitness in models with constant population size, making the equivalence possible. 

The fitness wave theory allows us to make analytical prediction about the properties of the antigenic wave. Let us start with its population size $N$, which is regulated by how fast the immune system catches up with the wave.
The immune turnover time $\tau$ in Eq.~\ref{eq:hvsn} is inversely proportional to $N$: the larger the population size, the faster immune memories are updated, increasing the immune pressure on current viral strains (lower $f_0$), and thus decreasing $N$.
As the moving wave reaches a stable moving state, its size $N$ becomes stable over time, giving the condition $(1/N)dN/dt=f_0=0$, which in turn constraints the ratio between the wave's size and speed:
\beq\label{eq:Noverv}
\frac{N}{v}=\frac{MN_h}{r}\left(R_0^{1/M}-1\right) = N_h s.
\eeq

\begin{figure}
\begin{center}
\includegraphics[width=\linewidth]{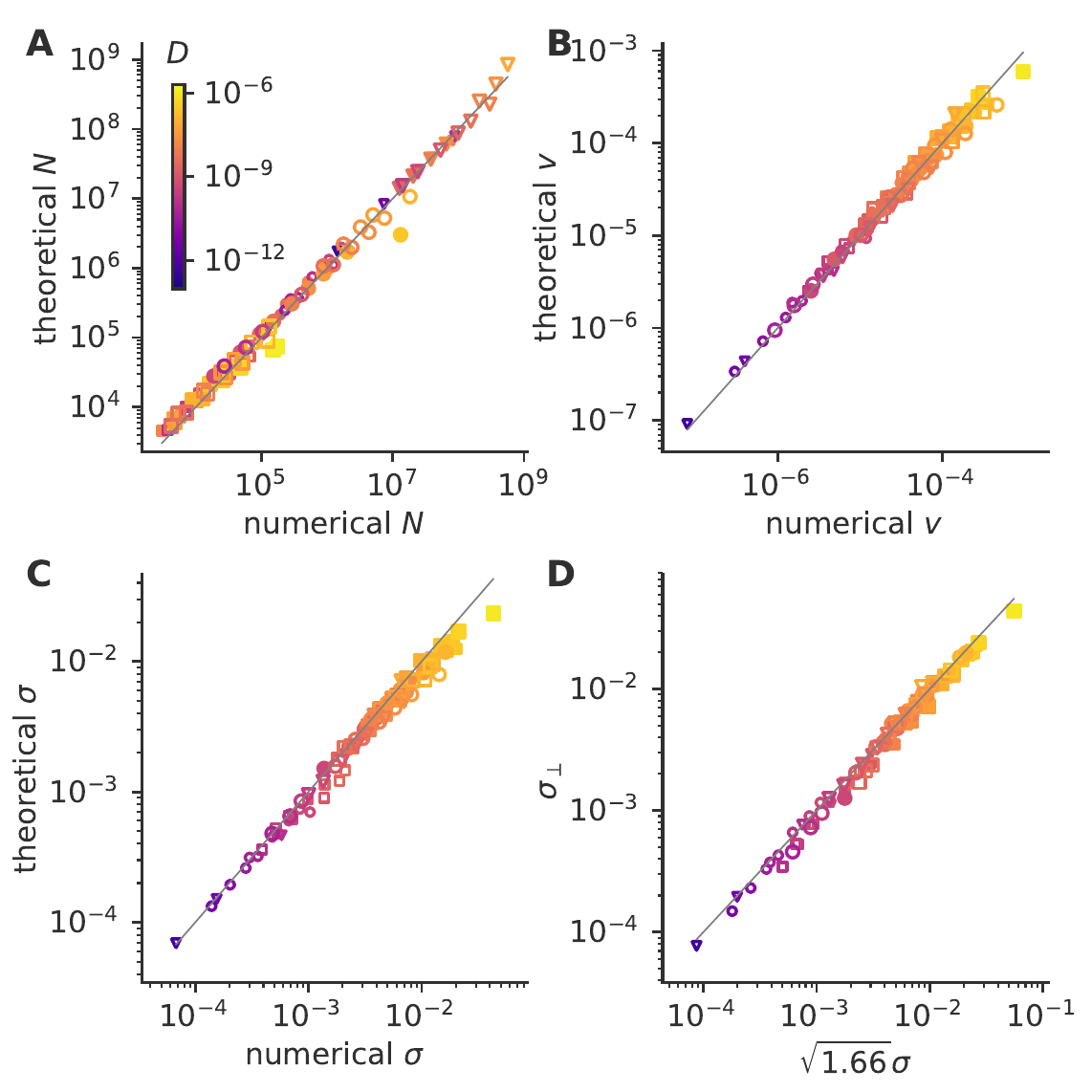}
\caption{{\bf Analytical prediction of wave properties.} Shown are the numerical versus analytical predictions for the wave's population size $N$  ({\bf A}), speed $v$ ({\bf B}), width $\sigma$ along the wave's direction of motion ({\bf C}), and width $\sigma_{\perp}$ in the direction perpendicular to motion ({\bf D}), with $d=2$ dimensions. Length are in units of the cross-reactivity range (so that $r=1$, with no loss of generality). Parameters: $N_h=10^8$ (squares), $10^{10}$ (circles), or $10^{12}$ (triangles); $\ln R_0=1$ (filled symbols) or $3$ (empty symbols); $M=1$ (small symbols) or $5$ (large symbols).}
\label{fig2}
\end{center}
\end{figure}

But the fitness wave theory predicts that the speed of the wave itself depends on the population size. The larger $N$, the more outliers at the nose of the fitness wave, and the further out they may jump in antigenic space, establishing fitter ancestors of the future population. This results in a fitness wave whose speed depends only weakly on population size and mutation rate (see \cite{Neher2013} and \app),
\beq\label{eq:speedf}
v_F \approx D_F^{2/3}\left[24\ln(N D_F^{1/3})\right]^{1/3},
\eeq
where $D_F=s^2D$ and $v_F=sv$ are the diffusivity and wave speed in fitness space, which are related to their counterparts in antigenic space through the scaling factor $s$. Replacing this scaling into Eq.~\ref{eq:speedf} yields a relation between antigenic speed and population size,
\beq\label{eq:speed}
v \approx {D}^{2/3}s^{1/3}\left[24\ln(N (Ds^2)^{1/3})\right]^{1/3},
\eeq
which closes the system of equations: using the definition of $s$ (Eq.~\ref{eq:s}), Eqs.~\ref{eq:Noverv} and \ref{eq:speed} completely determine $N$ and $v$ as a function of the model's parameters (through a transcendental equation, see \app).
We validated these theoretical predictions for $N$ and $v$ by comparing them to numerical simulations, which show good agreement over a wide range of parameters (Fig.~\ref{fig2}A-B). {\ch We note that the alternative fitness wave model of Desai and Fisher \cite{Desai2007b} predicts different scaling relations between speed and population size, including for an arbitrary distribution of fitness effects \cite{Good2012}. The major difference with our description is that we assume infinitesimal and reversible fitness effects. Relaxing that assumption to account for rare but strong mutational effects would affect Eq.~\ref{eq:speed}, but the dependence on $N$ would still be logarithmic at most.}

\subsection{Shape of the antigenic wave} 

The width $\sigma$ of the wave in the direction of motion is given by Fisher's theorem, which relates the rate of change of the average fitness to its variance in the population: $\partial_t f=\mathrm{Var}(f)$. In our description fitness and the antigenic dimension $x_1$ are linearly related with coefficient $s$, implying $s^2\sigma^2=sv$. The result of that prediction for $\sigma$ is validated against numerical simulations in Fig.~\ref{fig2}C.

The wave is led by an antigenic `nose' formed by few outlying strains of reduced cross-reactivity with the concurrent immune population, generating high fitness. These strains have phenotype $u_c =s\sigma^4/4D=v^2/(4Ds)$ and fitness $su_c$. They serve as founder strains from which the bulk of the future population will derive some time $\sim u_c/v=\sigma^2/4D$ later (see \app). As a result, two strains taken at random can trace back their most recent common ancestor to some average time $\<T_2\>=\alpha\sigma^2/2D$ in the past, where $\alpha\approx 1.66$ is a numerical factor estimated from simulations \cite{Neher2013}.

To explain the width $\sigma_{\perp}$ of the wave in the other phenotypic dimensions than that of motion ($x_{i>1}$), we note that in these directions evolution is neutral. Two strains taken at random in the bulk are expected to have drifted, or `diffused' in physical language, by an average squared displacement $\<\Delta x_i^2\>=2DT_2$ from their common ancestor, so that their mean squared distance is $4D\<T_2\>=2\alpha\sigma^2$ along $x_i$. If one assumes an approximately Gaussian wave of width $\sigma_{\perp}$, the mean square distance between two random strains along $x_i$ should be equal to $2\sigma_\perp^2$. Equating the two estimates yields $\sigma_\perp^2=\alpha \sigma^2$. Fig.~\ref{fig2}D checks the validity of this prediction against simulations.

Both longitudinal and transversal fluctuations in antigenic space are  instances of quantitative traits under interference selection generated by multiple small-effect mutations. The width of these traits is governed by the  common relation $\<\Delta x_i^2\>=2D\<T_2\>\sim \sigma^2$,  which expresses the effective neutrality of the underlying genetic mutations~\cite{Held2019}. This relation says that antigenic variations in all dimensions scale in the same way with the model parameters, and the wave should have an approximately spherical shape. Consistently, here we find a wave with a fixed ratio $\alpha \approx 1.66$ between transverse and longitudinal variations. This implies a slightly asymmetric shape (which may be non-universal and depend on the microscopic assumptions of our mutation model).

In what parameter regime is our theory valid? The fitness wave theory we built upon is meant to be valid in the large population size, $N\gg 1$. In addition, we assumed that the fitness landscape was locally linear across the wave. This approximation should be valid all the way up to the tip of wave, given by $u_c$, since this is where the selection of future founder strains happen. This condition translates into $u_c\ll r$, implying {\ch $D\ll r^2/\ln(N)^2$ (using $u_c=v^2/(4Ds)$ and Eqs.~\ref{eq:s},\ref{eq:speed})}, where $D$ is in antigenic unit squared per infection cycle. This result means that one infection cycle will not produce enough mutations for the virus to leave the cross-reactivity range. In that limit, another assumption is automatically fulfilled, namely that the width of the wave be small compared to the span of immune memory: $\sigma\ll v\tau$. Our simulations, which run in the regime of very slow effective diffusion ($D/r^2\lesssim 10^{-6}$) and have relatively large population sizes ($N\gtrsim 10^4$), satisfy these conditions. This explains the good agreement between analytics and numerics.

\begin{figure}
\begin{center}
\includegraphics[width=\linewidth]{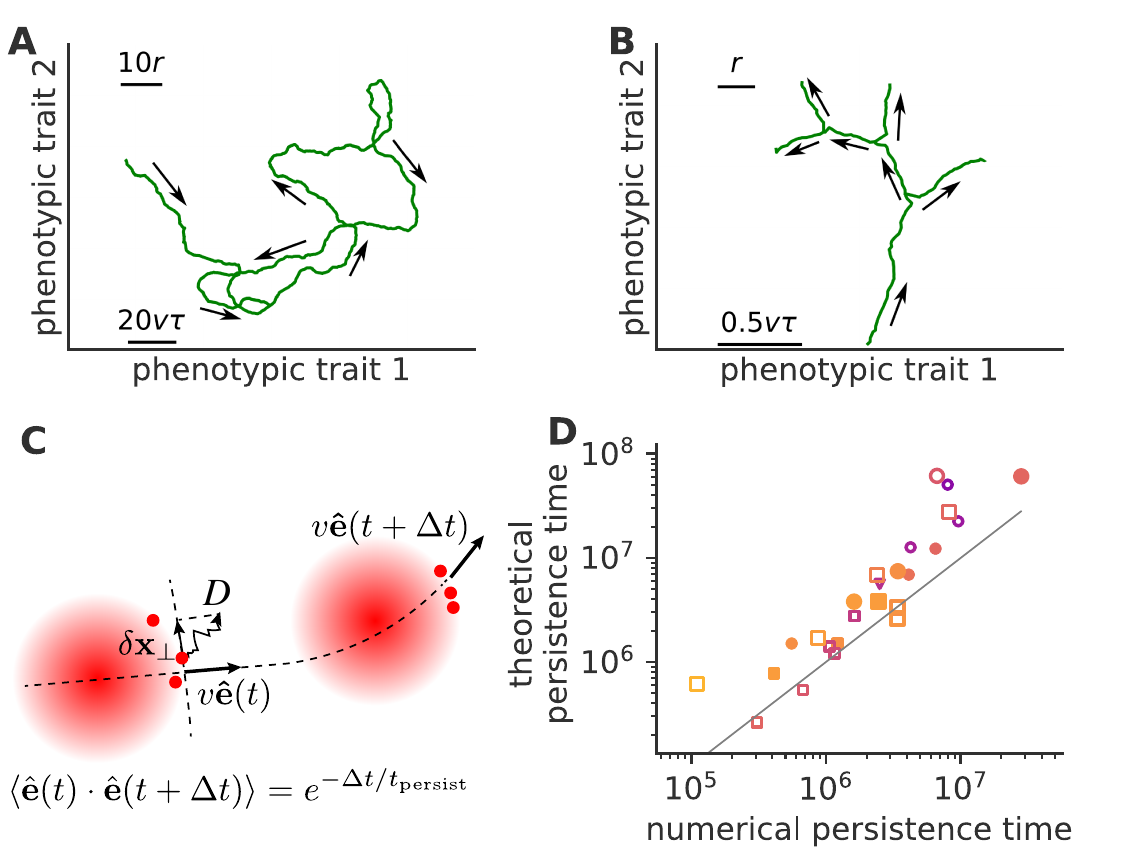}
\caption{{\bf Stochastic behaviour of the wave: diffusive motion, splits, and extinctions.}
  {\bf A.} The wave moves forward in antigenic space but is driven by its nose tip, which undergoes antigenic drift (diffusion) in directions perpendicular to its direction of motion. These fluctuations deviate that direction, resulting in effective angular diffusion.
  {\bf B.} When antigenic drift is large, the wave may randomly split into subpopulations, creating independent waves going in different directions. Each wave can also go extinct as size fluctuations bring it to 0.
  {\bf C.} Cartoon illustrating the wave's angular diffusion. Selection and drift combine to create a inertial random walk of persistence time $t_{\rm persist}$.
  {\bf D.} Analytical prediction (Eq.~\ref{eq:persist}) for the persistence time, versus estimates from simulations. Symbols and colors are the same as in Fig.~\ref{fig2}.
  }
\label{fig3}  
\end{center}
\end{figure}

\subsection{Equations of motion of the wave's position}
The wave solution allows for a simplified picture. The wave travels in the direction of the fitness gradient (or equivalent the gradient of immune coverage) with speed $v$ (Fig.~\ref{fig3}A). Occasionally the population splits into two separate waves that then travel away from each other and from their common ancestor (Fig.~\ref{fig3}B).
The tip of the wave's nose, which contains the high-fitness individual that will seed the future population, determines its future position in antigenic space. 
In the directions perpendicular to the fitness gradient, this position diffuses neutrally with coefficient $D$. This motivates us to write effective equations of motion for the mean position of the wave:
\begin{align}
\label{eq:eff1}
&\frac{d\bx}{dt}=-\left(v+\sqrt{2D_\parallel}{\xi}_\parallel(t)\right)\frac{\partial_\bx c}{| \partial_\bx c|}+\sqrt{2D}{\boldsymbol\xi}_\perp(t),\\
\label{eq:eff2}
&c(\bx,t)=\int_{-\infty}^t \frac{dt'}{\tau}e^{-\frac{t-t'}{\tau}-\frac{|\bx-\bx(t')|}{r}},
\end{align}
where ${\xi}_\parallel$ and ${\boldsymbol\xi}_\perp$ are Gaussian white noises in the directions along, and perpendicular to, the fitness gradient $\partial_\bx f/|\partial_\bx f|=-\partial_\bx c/|\partial_\bx c|$. $D_{\parallel}$ is an effective diffusivity in the direction of motion resulting from the fluctuations at the nose tip. These fluctuations are different than suggested by $D$, as they involve feedback mechanisms between the wave's speed $v$, size $N$, and advancement of the fitness nose $u_c$. In the following, we do not consider these fluctuations, and focus on perpendicular fluctuations instead.

\subsection{Angular diffusion and persistence of the antigenic wave}
In the description of Eqs.~\ref{eq:eff1}-\ref{eq:eff2}, the viral wave is pushed by immune protections left in its trail. The fitness gradient, and thus the direction of motion, points in the direction that is set by the wave's own path. This creates an inertial effect that stabilizes forward motion. On the other hand, fluctuations in perpendicular directions are expected to deviate the course of that motion, contributing to effective angular diffusion. To study this behaviour, we assume that motion is approximately straight in direction $x_1=vt$, and study small fluctuations in the perpendicular directions, $\bx_\perp=(x_2,\ldots,x_d)$, with $|\bx_\perp|\ll r$ (as illustrated in Fig.~\ref{fig3}C). Eqs~\eqref{eq:eff1}-\eqref{eq:eff2} simplify to (see \app):
\beq\label{eq:effperp}
\partial_t \bx_\perp(t)=\int_0^{+\infty} \frac{dt'}{T}\frac{\bx_\perp(t)-\bx_\perp(t-t')}{t'}e^{-t'/T}+\sqrt{2D}{\boldsymbol\xi}_\perp(t),
\eeq
where $T=(v/r+1/\tau)^{-1}=(r/v)R_0^{-1/M}$ is an effective memory timescale combining the host's actual immune memory, and the cross-reactivity with strains encountered in the past.

Eq.~\ref{eq:effperp} may be solved in Fourier space. Defining
$\tilde \bx_\perp(\omega)=\int_{-\infty}^{+\infty}dt e^{i\omega t}\bx_\perp(t)$,
it becomes:
\beq\label{eq:fourrier}
-i\omega \tilde \bx_\perp(\omega) \left(1+\frac{\ln(1-i\omega T)}{i\omega T}\right)= \sqrt{2D}\tilde {\boldsymbol\xi}_\perp(\omega).
\eeq
To understand the behaviour at long times $\gg T$, we expand at small $\omega$: $
-\omega^2 \bx_\perp(\omega) \approx \sqrt{8D}\tilde {\boldsymbol\xi}_\perp(\omega) /{T}$ or equivalently in the temporal domain $\partial_t^2 \bx_\perp \approx {\sqrt{8D}}{\boldsymbol\xi}_\perp (t) /{T}$.
This implies that the direction of motion, ${\hat{{\bf e}}}\sim \partial_\bx f/|\partial_\bx f|\sim\partial_t \bx/|\partial_t \bx|$, undergoes effective angular diffusion in the long run:
$\partial_t {\hat{{\bf e}}}={\sqrt{8D}}{\boldsymbol\xi}_\perp (t) /({vT}).$
The persistence time of that inertial motion,
\beq\label{eq:persist}
t_{\rm persist}=\frac{v^2T^2}{4D}=\frac{r^2}{4D}R_0^{-2/M},
\eeq
does not depend explicitly on speed, population size, or the dimension of antigenic space. However, a larger diffusivity implies larger $N$ and $v$ while reducing the persistence time. Likewise, a larger reproduction number $R_0$ or smaller memory capacity $M$ speeds up the wave and increases its size, but also reduces its persistence time. This implies that, for a fixed number of hosts $N_h$, larger epidemic waves not only move faster across antigenic space, but also change course faster.

This persistence time scales as the time it would take a single virus drifting neutrally to escape the cross-reactivity range, $r^2/D$. For comparison, the much shorter timescale for a {\em population} of viruses to escape from the cross-reactivity range $r$, 
\beq
t_{\rm escape} = \frac{r}{v} = TR_0^{1/M}= \frac{N_h M}{N}(R_0^{1/M}-1), 
\eeq
scales with the inverse incidence rate $N_h/N$.
This is consistent with the whole population having been infected at least one every $\sim N_h/N$ infection cycles.
This separation of time scales is consistent with the observation that evolution in the transverse directions is driven by neutral drift, which is much slower than adaptive evolution in the longitudinal direction. Both $t_{\rm persist}$ and $t_{\rm escape}$ are longer than the coalescence time of the viral population, $u_c/v\sim \sigma^2/4D$, since they reflect long-term memory from the immune system. However, while $t_{\rm escape}\sim N_h/N$ is related to the re-infection period and is thus bounded by the hosts' immune memory (itself bounded by their lifetime, which we do not consider), $t_{\rm persist}$ can be  longer than that. This is possible thanks to inertial effects, which are allowed by the high-order dynamics of Eq.~\ref{eq:effperp} generated by the immune system. This very much like when, in mechanics, a massive object set in motion in a given direction will keep that direction without the need for an external force to maintain it.

The high-frequency behaviour of \eqref{eq:fourrier} has a logarithmic divergence, meaning that the total power of ${\hat{{\bf e}}}$ is infinite unless we impose a (ultraviolet) cutoff. Such a regularization emerges from the fine structure of the wave. While the motion of the wave is driven by its nose tip, the immune pressure only extends back to the recent past of the bulk of the distribution, which stands at a distance $u_c$ away from the nose. In other words, there is a lag (and thus an gap $u_c$ in antigenic space) between the most innovative variants that drive viral evolution, and the majority of currently circulating variants which drive host immunity.
Mathematically, this implies that the domain of integration of the first term in the right-hand side of \eqref{eq:effperp} should start at $t_c=u_c/v$, which regularizes the divergence. A more careful analysis provided in the {\app} shows that this regularization does not affect the long-term diffusive behaviour of the wave.

\subsection{Canalization, speciations, and predictability of antigenic evolution}

We now examine how deflections of the wave in the transverse direction determines the predictability and stability of the viral quasi-species.
Assuming $t\gg T$, angular diffusion causes motion to be deflected as (see \app)
$\<x_{\perp}^2\>=\frac{8(d-1)D}{3T^2}t^3.$
Crucially, this deflection depends on the dimension of the antigenic space, because the displacement acts additively in each of the transversal coordinates. Higher dimension means more deviation from the predictable course of the wave, and thus less predictability. We can define a predictability time scale
\beq
t_{\rm predict}\sim [8(d-1)/3]^{-1/3}T^{2/3}(r^2/D)^{1/3},
\eeq
which is the time it takes for prediction errors to become of the order of the cross-reactivity range. 
In low dimensions, this time scales as a weighted geometric mean between $t_{\rm escape}\sim T$ and $t_{\rm persist}\sim r^2/D$. However, at high dimensions $t_{\rm predict}$ may be significantly reduced, causing loss of predictability even below $t_{\rm escape}$. {\ch The prediction timescale is distinct from the previously discussed persistence time: $t_{\rm predict}$ involves the integrated displacement in the transversal direction, while $t_{\rm persist}$ quantifies the diffusion of the tangent velocity vector. Thus, $t_{\rm predict}$ may be interpreted as quantifying the predictability of the actual location of the next viral population in antigenic space, while $t_{\rm persist}$ gives the predictability of the general direction of evolution, which changes more slowly. Therefore, the persistence time is both harder to extract from data and less relevant for actionable predictions.}

To get a sense of numbers, we can compare our results with epidemiological data, taking the evolution of influenza as an example, with an infection cycle time of 3 days. It is assumed that individuals lose immunity to the circulating strain of the flu within $\sim 5$ years $\sim 500$ cycles, meaning that the wave would travel a distance $r$ in $t=500$, i.e. $v/r\sim 2\cdot 10^{-3}$. For instance, with $N_h=10^9$-$10^{10}$, $R_0=2$, and $M=1$, {\ch we may choose $D/r^2=3\cdot 10^{-6}$ to get a speed of the same order, $v/r\sim 1.3\cdot 10^{-3}$,} and $t_{\rm persist}\sim 2\cdot 10^4\sim 200$ years. By contrast, the predictability timescale $t_{\rm predict}$ is much shorter and depends on dimension, albeit slowly, ranging from $\sim 20$ years for $d=2$ to about 2 years for $d=1000$.
We stress that these numbers are obtained by scaling laws, and should not be taken as precise quantitative predictions. 

\begin{figure}
\begin{center}
  \noindent\includegraphics[width=\linewidth]{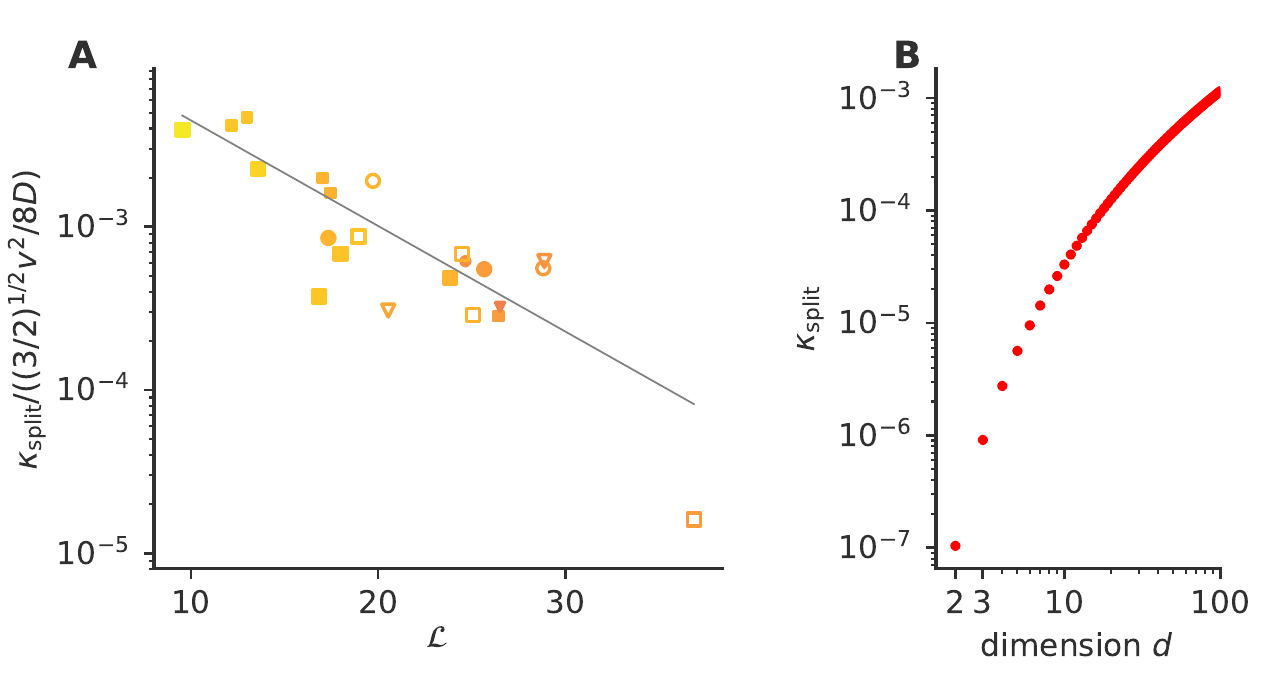}
\caption{{\bf Rate of speciation. A.} Rescaled rate of splitting events, defined as the emergence of two substrains at distance $\Delta x_0=0.1 r$ from each other in antigenic space, meaning that they are becoming antigenically independent. The predicted scaling, $k_{\rm split}\sim (v^2/D)e^{-\mathcal{L}}$, as well as the definition of the collective variable $\mathcal{L}$ as a function of the model parameters, are given by Eq.~\ref{eq:split}. The line shows a linear fit of the logarithm of the ordinate. {\ch {\bf B.} Predicted rate of splitting as a function of the dimension $d$, for $R_0=2$, $M=1$, $N_h=10^9$, and $D/r^2=3\cdot 10^{-6}$, with $\Delta x_0=r$.}
  }
\label{fig4}  
\end{center}
\end{figure}

Large deflections may also cause speciations, or splits, which occur when two substrains co-exist long enough to become independent from the immune standpoint. This happens when two sub-lineages see the difference of their transverse positions $\Delta\bx_{\perp}$ become larger than $\Delta x_0\sim r$, within some limited period given by the coalescence time. We estimated the rate of such splitting events using a saddle-point approximation (see \app):
\beq\label{eq:split}
k_{\rm split}\approx
\sqrt{\frac{3}{8}}\frac{v^2}{4D}e^{-\mathcal{L}},\quad \mathcal{L}=\alpha{\left(\frac{s^3R_0^{-2/M}D^2r^4}{(d-1)v^5}\right)}^{1/4}
\eeq
with $\alpha$ some numerical factor. Simulations confirmed the validity of this scaling (Fig.~\ref{fig4}a).

The splitting rate grows with the dimension {\ch (Fig.~\ref{fig4}b)}, consistent with the intuition that departure from canalized evolution is easier when more directions of escape are available. Splitting events are expected to strongly affect our ability to predict the future course of the wave. However, the rarity of such events (exponential scaling of $k_{\rm split}$) means that they will have a lower impact on predictability than deflections. These results provide a theoretical and quantitative basis from which to assess the effect of dimension on predictability, and possibly estimate $d$ from antigenic time course data of real viral populations.

\section{Discussion}
In this work, we have developed an analytical theory for studying antigenic waves of viral evolution in response to immune pressure. We showed that predictabilty is limited by two features of antigenic evolution, transversal diffusion and lineage speciations of the antigenic wave, both of which explicitly depend on the dimensionality of antigenic space.

{\ch To derive these results, we explicitly embedded the antigenic phenotype in a $d$-dimensional Euclidean space. This description is different from previous work that considered one- \cite{Rouzine2018} or infinite-dimensional antigenic spaces \cite{Yan2019}. It allows for} the possibility of compensatory mutations, and makes it easier to compare results with empirical studies of viral evolution projected onto low-dimensional spaces \cite{Smith2004,Bedford2014}. Unlike these studies, however, our work does not address the question how an effective dimension of antigenic space arises from the molecular architecture of immune interactions. Rather, we focused on the implications of the dimensionality of antigenic space for phenotypic evolution and its predictability.

Our results suggest a hierarchy of time scales for viral evolution. The shortest is the coalescence time $\<T_2\>$, which determines population turnover. Then comes $t_{\rm escape}$, which is the time it takes the viral population to escape immunity elicited at a previous time point. The longest timescale is the persistence time $t_{\rm persist}$, which governs the angular diffusion of the wave's direction, {\ch but has no bearing on the prediction of the actual position of the dominant strain in antigenic space}. That time scale is due to inertial effects. {\ch It does not} rely directly on the hosts' immune memories, and may thus exceed their individual lifetimes.  Finally, the prediction timescale $t_{\rm predict}$, beyond which prediction accuracy falls below the resolution of cross-reactivity, scales between $t_{\rm escape}$ and $t_{\rm persist}$ at low dimensions. {\ch This time scales measures the predictability of transversal fluctuations, and is thus the most relevant for actual predictions of future dominant strains in antigenic space.
Importantly,} it decreases with the dimension of the antigenic space, and may become arbitrarily low at very high dimensions. {\ch The fact that the evolution of influenza strains are hard to predict beyond a year suggests that the effective dimension may indeed be large.}

{\ch Our solution builds on the fitness wave solution for a diffusion model of mutation effects \cite{Cohen2005,Neher2013}. It implies a particular dependence of the wave's speed on the population size, Eq.~\ref{eq:speed}. General distribution of non-infinitesimal mutational effects, such as considered in \cite{Good2012}, would yield different expressions for the speed. However, we expect most of our other results to hold---in particular, all expression that do not carry an explicit logarithmic dependence on $N$, as well the effective equations of motion for the wave. Our results strongly rely on the assumption of a homogeneous, isotropic antigenic space. We expect our results to be affected by anisotropies (e.g. in the mutational or the cross-reactivity Kernels), or by structure in the {\em intrinsic} fitness landscape (i.e. not linked to immunity). Such structure may funnel the wave in preferred directions, hinder it, or favor its splitting. Generally, the local geometry and metric of the space is expected to determine the evolutionary behaviour.  For instance, Yan et al. \cite{Yan2019} assumed a Hamming distance metric in an effectively infinite antigenic space, meaning that any mutation is both an escape mutation and a candidate for a lineage split. By contrast, in our geometry, escape happens only in the direction of the wave, while splits originate from mutations perpendicular to that direction, due to the choice of a Euclidean metric. While our results emphasize the role of the effective dimension $d$, studying other geometrical effects is an interesting topic for future work.
}

{\ch 
  Despite these caveats, it is interesting to ask whether the effective antigenic dimension $d$ can be extracted from data. A possible scheme for doing so starts by inferring the effective model parameters. $R_0$ may be estimated from exponential epidemic growth in a susceptible population. Dependence of key quantities on $M$ such as $s$ is weak. $M$ may be assumed to be of the order of the number of antigenically distinct infections encountered during a host's lifetime, $\sim 4$-$6$ (every 15 years). $D/r^2$ may be inferred from $v/r$, which can be estimated from cross-immunity assays or from the incidence rate $N/N_h$. Alternatively, since $D/r^2$ is the inverse time it takes for mutations to neutrally evade immunity, it could be estimated directly from genomic data by computing the time for unselected mutations (whose rate is inferred from synonymous mutations) to affect antigenic sites. Interestingly, if $v/r$ and $D/r^2$ can be inferred independently, predictions about the wave's shape, width, angular diffusion and splitting do not depend on the particular choice of fitness wave theory.
  Assuming that all these parameters are known, the splitting rate, which depends sensitively on $d$ (Fig.~\ref{fig4}b), could be used to infer an effective dimension. Since splitting is rare and may not be observed in practice, one could define instead partial splits, where a sublineage diverges an antigenic distance $\Delta x_0<r$ from the main lineage, for which the same scaling as Eq.~\ref{eq:split} holds (see \app).
  Alternatively, our results could be used to check the consistency of dimensionality-reduction schemes based on serological assays \cite{Smith2004,Bedford2014,Fonville2014}, by testing our predicted relations between the speed of the wave, its width and length, and angular diffusion properties, and ask what choice of dimension best agrees with our theory.
  }

Our framework should be applicable to general host-pathogens systems. For instance, co-evolution between viral phages and bacteria protected by the CRISPR-Cas system \cite{Westra2019} is governed by the same principles of escape and adaptation as vertebrate immunity. Even more generally, our theory (Eqs.~\ref{dndt},\ref{dhdt}) may be relevant to the coupled dynamics of predators and preys interacting in space (geographical or phenotypic), opening potential avenues for experimental tests of these theories in synthetic microbial systems. {\ch Given the current context of the global SARS-CoV-2 pandemic, it is natural to ask whether our results could be applicable to predict its evolution. While our theory describes the long-term co-evolution of viral strains with the hosts' immune systems, in which most hosts have been exposed to at least on strain of the virus, SARS-CoV-2 is still in a phase of growth, and has not exhausted the reservoir of susceptible hosts. As the situation develops, it will be interesting to see whether its future evolution follows a Red Queen type of evolution like influenza, goes extinct, or splits into many antigenically independent sublineages. While our model may shed light on these questions, fine microscopic details such as geographical and population structure impose additional challenges for predictions.}

\section{Methods}
We simulated discrete population dynamics of infected hosts $n(\bx,t))$ and immune protections $n_h(\bx,t)\equiv N_hh(\bx,t)$ (all integers) on a $2D$ square lattice with lattice size $\Delta x$ ranging from $10^{-5}r$ to $0.1r$.
Each time step corresponds to a single infection cycle, $\Delta t=1$. At each time step: {\em (1)} viral fitness $f$ is computed at each occupied lattice site from the immune coverage Eq.~\ref{fit_def}; {\em (2)} viruses at each occupied lattice site are grown according to their fitness, $n(\bx,t+1)\sim \mathrm{Poisson}[(1+f\Delta t)n(\bx,t)]$; {\em (3)} viruses are mutated by jumping to nearby sites on the lattice; {\em (4)} the immune system is updated according to a discrete version of Eq.~\ref{dhdt}, by implementing $n_h(\bx,t+1)=n_h(\bx,t)+n(\bx,t)$ and then removing $N(t)$ protections at random (so that $N_h$ remains constant).

To implement {\em (1)}, we used a combination of exact computation of  Eq.~\ref{fit_def} and approximate methods, including one based on non-homogeneous fast Fourier transforms \cite{keiner2009using,potts2004fast}. Details are given in the {\app}.

To implement {\em (3)}, we drew the number of mutants at each occupied site from a binomial distribution $\mathrm{Binomial}(n(\bx,t) ,1-e^{-\mu\Delta t})$. The number of new mutations $m$ affecting each of these mutants  is drawn from a Poisson distribution of mean $\mu \Delta t$ conditioned on having at least $1$ mutation.
The new location of each mutant is drawn as $\bx+\delta \bx$, with $\delta\bx=\mathrm{round}(\sum_{i=1}^m{\boldsymbol\epsilon}_i)$
 (rounding is applied to each dimension), where ${\boldsymbol\epsilon}_i$ is a vector of random orientation and modulus drawn from a Gamma distribution of mean
$\delta\sim 2 \Delta x$
and shape parameter $20$. This distribution was chosen so as to maximize the number of non-zero jumps while maintaining isotropy. We then define $D=\mu \<\delta x_1^2\>/2$.

To find the wave solution more rapidly, the viral population was initialized as a Gaussian distribution centered at $(0,0)$ with size $N$ and width $\sigma$ in all dimensions, to which $0.1\%$ additional viruses are randomly added within the interval $(0;u_c)$ along $x_1$ ($N$, $\sigma$, and $u_c$ being all given by the theory prediction). Immune protections are placed according to Eq.~\ref{happrox}. The first 20,000 time steps serve to reach steady state and are discarded from the analysis. When a population extinction ($N=0$) or explosion ($N=N_h/2$) occurs, the simulation is resumed at an earlier checkpoint to avoid re-equilibrating. Simulations are ended after $5\cdot 10^6$ steps or after 20 consecutive extinctions or explosions from the same checkpoint.


In order to analyze the organization of viruses in phenotypic space,  
we save snapshots of the simulation at regular time intervals.
For each saved snapshot we take  all the coordinates with $n>0$ and then cluster them into separate lineages through the python scikit-learn DBSCAN algorithm~\cite{scikitlearn}~\cite{dbscan} 
with the minimal number of samples ${\rm min\_samples} = 10$. The $\epsilon$ parameter defines the maximum distance between two samples that are considered to be in the neighborhood of each other. We perform the clustering for different values of $\epsilon$ and select the value that minimizes the variance of the 10th nearest neighbor distance. Clustering results are not sensitive to this choice. 
This preliminary clustering step is refined by merging clusters if  their centroids  are closer than the sum of the maximum distances of all the points in each cluster from the corresponding centroid. 

From the clustered lineages we can easily obtain a series of related observables, such as its speed $v$ obtained as the derivative of the center's position. The width of the lineage profile in the direction of motion $\sigma$ as well as in the perpendicular direction $\sigma_{\perp}$ are obtained by taking the standard deviaton of the desired component of the distances of all the lineage viruses from the lineage centroid. Reported numbers are time averages of these observables. 
We can track their separate trajectories in antigenic space. A split of a lineage into two new lineages is defined when two clusters are detected where previously there was one, and their distance is larger than $\Delta x_0$, {\ch the chosen threshold for calling a split}.

To estimate the persistence time, we first subsample the trajectory so that the distance between consecutive points is bigger than $6 (\<\sigma\> + {\rm std}(\sigma))$ so that fast fluctuations in the population size do not affect the inference. We take the resulting trajectory angles and smooth them with a sliding window of $5$.
Then we divide the trajectory into subsegments, and compute the angles mean squared displacement (MSD) over all lineages and all subsegments.  We consider time lags only bigger than twice the typical smoothing time, and if the MSD trace is long enough we also require the time lag to be bigger than $2T$. Finally we only keep time lag bins with at least $10$ datapoints. We fit the resulting time series to a linear function $a x + b$, and get the persistence time as $\frac{2}{a}$. We compute the reduced $\chi^2$ as a goodness-of-fit score. Results are shown for simulations that had enough statistics to perform the fit, lasted at least $10^5$ cycles, and had a reduced $\chi^2$ below $3$.

{\bf Acknowledgements.}
The study was supported by the European Research Council COG 724208
and ANR-19-CE45-0018 ``RESP-REP'' from the Agence Nationale de la Recherche and DFG grant CRC 1310
``Predictability in Evolution''.

\bibliographystyle{pnas}

\onecolumngrid

\appendix

\renewcommand{\thefigure}{S\arabic{figure}}
\setcounter{figure}{0}

\section{Fitness wave theory}
We decompose the density of viral strains according to the main
direction of the wave $x_1$:
\beq
n(\bx,t)=n_1(x_1,t)\phi(x_2,\ldots,x_d),
\eeq
where $\phi$ is normalized to 1. Projecting and linearizing Eq.~\ref{dndt} of the main text yields:
\beq
\frac{\partial n_1(x_1,t)}{\partial
  t}=s(x_1-vt)  n_1(x_1,t)+D\frac{\partial^2 n_1}{\partial x_1^2}+\sqrt{n_1(\bx,t)}\eta_1(x_1,t),
\eeq
with $s$ defined by \eqref{eq:s}, and $\eta_1=\int dx_2\cdot
dx_d\,\eta(\bx,t)$, so that $\<\eta_1(x_1,t)\eta_1(x'_1,t')\>=\delta(x_1-x'_1)\delta(t-t')$. The change of variable $\tilde x_1=sx_1$, $\tilde
v=sv$, $\tilde n_1=s^{-d}n_1$, yields the traveling wave equation of
Ref.~\cite{Neher2013}:
\beq
\frac{\partial \tilde n_1(\tilde x_1,t)}{\partial
  t}=(\tilde x_1-\tilde vt)\tilde n_1(x_1,t)+\tilde D\frac{\partial^2
  \tilde n_1}{\partial \tilde x_1^2}+\sqrt{\tilde
  n_1(\bx,t)}\tilde\eta_1(\tilde x_1,t),
\eeq
with $\tilde D=Ds^2$ and $\<\tilde\eta_1(\tilde x_1,t)\tilde\eta_1(\tilde x'_1,t')\>=\delta(\tilde x_1-\tilde x'_1)\delta(t-t')$.

Note that this continuous description differs from that used in
Ref. \cite{Yan2019}, which also describes a fitness wave in antigenic
space. Their approach relies on a discrete evolutionary model where each mutation confers a fixed fitness advantage, as described by the fitness wave solution of Desai and Fisher \cite{Desai2007b}.

Applying the formulas of {\ch the diffusive theory \cite{Neher2013}} yields in the limit of large populations:
\beq
s^2\sigma^2 \approx \tilde D^{2/3}(24\ln(N\tilde D^{1/3}))^{1/3},
\eeq
or
\beq
\sigma \approx (D/s)^{1/3}(24\ln(N (Ds^2)^{1/3}))^{1/6},
\eeq
and
\beq
v \approx {D}^{2/3}s^{1/3}(24\ln(N (Ds^2)^{1/3}))^{1/3}.
\eeq
The fittest in the population is ahead of the bulk by
$u_c=s\sigma^4/4D$ in phenotypic space, with
\beq
u_c\approx \frac{1}{4}(D/s)^{1/3}(24\ln(N (Ds^2)^{1/3}))^{2/3}
\eeq

Plugging in \eqref{eq:s} yields:
\beq
\sigma=\left(\frac{D r}{M ( R_0^{1/M}- 1)} \right)^{1/3}\left(24\ln\left(N D^{1/3} \left(\frac{M ( R_0^{1/M} - 1)}{r} \right)^{2/3}\right)\right)^{1/6},
\eeq
\beq
v={D}^{2/3}\left(\frac{M ( R_0^{1/M}- 1)}{r} \right)^{1/3} \left(24\ln\left(N D^{1/3} \left(\frac{M (  R_0^{1/M}- 1)}{r} \right)^{2/3}\right)\right)^{1/3},
\eeq
\beq
u_c\sim \frac{1}{4}\left(\frac{D r}{M ( R_0^{1/M} - 1)} \right)^{1/3}\left(24\ln\left(N D^{1/3} \left(\frac{M ( R_0^{1/M} - 1)}{r} \right)^{2/3}\right)\right)^{2/3}.
\eeq
From the stationarity condition \eqref{eq:Noverv} we obtain
a self-consistent equation for $N$:
\beq
 \frac{N}{N_h} = \frac{M}{\tau}  = s v = {D}^{2/3}\left(\frac{M ( R_0^{1/M} - 1)}{r} \right)^{4/3} \left(24\ln\left(N D^{1/3} \left(\frac{M ( R_0^{1/M} - 1)}{r} \right)^{2/3}\right)\right)^{1/3}.
\eeq

The condition $u_c\ll r$, implies that $r$ scales with $N$ faster than
$u_c$, $r  \gg \ln(N)$. We also want $\sigma \ll v\tau$, therefore $r  \gg \frac{M ( R_0^{1/M}
- 1)}{M^{3/2}}  \ln(N)^{1/4}$, which is automatically satisfied by the
previous condition.

\section{Fluctuations in the direction perpendicular to motion}
The dynamics of the wave in the directions that are orthogonal to $x_1$
is governed by the projection of \eqref{eq:eff1} onto $\bx_\perp=(x_2,\ldots,x_d)$:
\beq\label{eq:effproj}
\partial_t \bx_\perp=-v \frac{\partial_{\bx_\perp} c}{|\partial_\bx
c|}+\sqrt{2D}{\boldsymbol \xi}_\perp.
\eeq
From \eqref{eq:eff2} we have
\beq
c(\bx,t)\approx \int_{-\infty}^t
\frac{dt'}{\tau}e^{-(t-t)'/\tau-\sqrt{(x_1-vt')^2+(\bx_\perp(t)-\bx_\perp(t'))^2}/r}.
\eeq
Taking the derivative along $\bx_\perp$ yields:
\beq
\begin{split}
\partial_{\bx_\perp}c|_{x_1=vt}&\approx\frac{1}{r}\int_{-\infty}^t
\frac{dt'}{\tau}\frac{-(\bx_\perp(t)-\bx_\perp(t'))}{\sqrt{(vt-vt')^2+(\bx_\perp(t)-\bx_\perp(t'))^2}}e^{-(t-t)'/\tau-\sqrt{(vt-vt')^2+(\bx_\perp(t)-\bx_\perp(t'))^2}/r}\\
& \approx -\frac{1}{rv\tau}\int_{-\infty}^t dt' \frac{\bx_\perp(t)-\bx_\perp(t')}{t-t'}e^{-(t-t')(1/\tau+v/r)},
\end{split}
\eeq
where we assumed $|\bx_\perp|\ll r,\tau/v$. This derivative is small
compared to the gradient along the $x_1$, so that we may approximate $|\partial_\bx c|\approx |\partial_{x_1}c|= (r+v\tau)^{-1}$.

Replacing into \eqref{eq:effproj},
we obtain:
\beq
\partial_t \bx_\perp\approx \int_{-\infty}^t \frac{dt'}{T}
\frac{\bx_\perp(t)-\bx_\perp(t')}{t-t'}e^{-(t-t')/T}+\sqrt{2D}{\boldsymbol
\xi}_\perp,
\eeq
with $1/T=1/\tau+v/r$. Using integration by part, this equation can be rewritten as an
auto-regressive process on $\partial_t\bx_\perp$:
\beq\label{eq:effsi}
\partial_t \bx_\perp=\int_0^{\infty} \frac{dt'}{T}E_1(t'/T)\partial_t \bx_\perp(t-t')+\sqrt{2D}{\boldsymbol
\xi}_\perp (t)
\eeq
where $E_1(x)\equiv\int_x^\infty dx'\,e^{-x'}/x'$ has the property
$\int_0^\infty dx\, E_1(x)=1$. Computing the Fourier transform of
$E_1(t/T)/T$, which is given by $-\frac{1}{i\omega
T}\ln(1-i\omega T)$ and using the rule of convolution in Fourier
space yields \eqref{eq:fourrier}. Eq.~\ref{eq:effsi} can be
re-written in terms of the direction of motion ${\hat{{\bf e}}}=\partial_t \bx/|\partial_t \bx|\approx \partial_t \bx/v$:
\beq
{\hat{{\bf e}}}(t)=\int_0^{\infty} \frac{dt'}{T}E_1(t'/T) {\hat{{\bf e}}}_\perp(t-t')+\frac{\sqrt{2D}}{v}{\boldsymbol
\xi}_\perp (t)
\eeq

Focusing on long-term behaviour yields the angular diffusion equation, $\partial_t {{\hat{{\bf e}}}}={\sqrt{8D}}{\boldsymbol\xi}_\perp (t) /({vT})$, for the direction of motion ${\hat{{\bf e}}}$. The two-point function of ${\hat{{\bf e}}}$ follows the equation
\beq
\partial_t \arccos \left[{\hat{{\bf e}}}(t_0) {\hat{{\bf e}}}(t_0+t)\right] = \frac{\sqrt{8D}}{vT} \eta(t),
\eeq
where $\eta(t)$ is a unit Gaussian white noise, leading to:
\beq
\<{\hat{{\bf e}}}(t) {\hat{{\bf e}}}(t+\Delta t)\>=e^{-\Delta t/t_{\rm persist}},
\eeq
with $t_{\rm persist}=v^2T^2/(4D)$ is defined as the persistence
time. Going along the curviline coordinate that follows the trajectory with
speed $v$, we obtain a persistence length of $vt_{\rm persist}$.

The logarithmic divergence at high frequencies in
\eqref{eq:fourrier}, which is also apparent in the logarithmic
divergence in the temporal domain at small $t$ in the auto-regressive
Kernel $E_1(t/T)/T$. This divergence may be regularized by realizing
that there is a lag $u_c/v$ between the nose of the wave, which drives the
behaviour of the wave, and its bulk. This implies that the integral
over the past trajectory encoding the immune memory extends only up
$t-u_c/v$ in the past:
\beq
\partial_t \bx_\perp\approx \int_{-\infty}^{t-u_c/v} \frac{dt'}{T}
\frac{\bx_\perp(t)-\bx_\perp(t')}{t-t'}e^{-(t-u_c/v-t')/T}+\sqrt{2D}{\boldsymbol
  \xi}_\perp,
\eeq
or after integration by parts:
\beq
\partial_t \bx_\perp=e^\epsilon \left[E_1(\epsilon)(\bx_\perp (t)-\bx_\perp (\bx_\perp-T\epsilon))+\int_{\epsilon T}^{\infty} \frac{dt'}{T}E_1(t'/T)\partial_t \bx_\perp (t-t')\right]+\sqrt{2D}{\boldsymbol
  \xi}_\perp,
\eeq
with $\epsilon=t_c/T\ll 1$.

In Fourier space this reads:
\beq
-i\omega \tilde \bx_\perp = K(-i\omega t)\tilde \bx_\perp +\sqrt{2D}\tilde {\boldsymbol
  \xi}_\perp
\eeq
where now the $K$ is the Laplace transform of the operator:
\beq
K(z)=e^\epsilon\left[E_1(\epsilon)-E_1(\epsilon(1+z))\right]
\eeq

Since $E_1(z)$ goes to $0$ for large $z$,
$(z-K(z))^{-1}$ goes as $1/z$ for $z\to \infty$ (small time scales), which
means that at high frequency fluctuations of $\partial_t \bx_\perp$
(and thus of the direction of motion ${\hat{{\bf e}}}$) track those of ${\boldsymbol
  \xi}_\perp$.

Since
$E_1(x)\sim -\gamma-\ln(x)+x$ at small $x$, then
$E_1(\epsilon)-E_1(\epsilon(1+z))\approx \ln(1+z)-\epsilon z$ for
moderate $z$ and small $\epsilon$, and
$K(z)\approx z-z^2/2+O(z^2\epsilon)$, so that $(z-K(z))^{-1}\sim
2/z^2$. We thus recover that at long time scales $\partial_t \bx_\perp$ diffuses with diffusivity
$4D/T^2$.

\section{Rate of speciation}

A speciation, or split, occurs when two strains starting at the tip of the nose
of the fitness wave, and continuing through their progenies, co-exist
long enough for them to become independent from the viewpoint of the
immune pressure. This happens when their distance in the $\bx_{\perp}$
direction become larger than some threshold scaling with the
cross-reactivity range, $\Delta x_0\sim r$.

Assuming $t\gg T$, angular diffusion causes a strain to bend from the
main direction of the wave as:
\beq
\partial_t^2 \bx_{\perp}=\frac{\sqrt{8D}}{T}{\boldsymbol\xi}_\perp (t).
\eeq
After integration, the expected deviation reads:
\beq
\<x_{\perp}^2\>=\frac{8(d-1)D}{3T^2}t^3
\eeq
assuming $x_\perp(t=0)=0$. Now if there are two strains $a$ and $b$ co-existing,
their divergence in the perpendicular direction is Gaussian
distributed with:
\beq
\<\Delta x^2\>=\<(x_{\perp a}-x_{\perp b})^2\>=\frac{16(d-1)D}{3T^2}t^3.
\eeq
Two strains are expected to co-exist at the leading edge for a time
$t$ before one of them gets absorbed into the bulk and goes
extinct. The expected time for that scales as $\sim\tau_{\rm
  sw}=u_c/v=\sigma^2/4D=v/4Ds$. Assuming splitting events are rare,
they occur when two co-existing strains both survive for an unusually
long time. The distribution of such rare events is asymptotically given by the
probability density function $P(t_{\rm coexist}>t)=e^{-t/\tau_{\rm
    sw}}$. The probability that the two strains have drifted by at
least $r$ before that happens is then given by:
\beq
P(\Delta x>\Delta x_0;{\rm coexist})=\int_0^{+\infty} \frac{dt}{\tau_{\rm
    sw}}e^{-t/\tau_{\rm sw}}\int_{\Delta x_0}^{+\infty} \frac{d\Delta x}{\sqrt{2\pi 16(d-1)Dt^3/(3T^2)}}\exp\left(-\frac{\Delta
      x^2}{32(d-1)Dt^3/(3T^2)}\right).
\eeq
Since we assume that this event is rare, $P(\Delta x>\Delta x_0; {\rm coexist})\ll 1$, we make a saddle-point approximation (Laplace
method) in the $t$ variable. We look for the maximum of
  \beq
  \mathcal{L}(t,\Delta x)=\frac{t}{\tau_{\rm sw}}+\frac{\Delta x^2}{32(d-1)Dt^3/(3T^2)},
 \eeq
    with respect to $t$, $\partial_t \mathcal{L}=0$, which gives:
    \beq
t^*= \frac{\sqrt{3}}{2^{5/4}} {\left(\frac{T^2\tau_{\rm sw}\Delta
      x^2}{(d-1)D}\right)}^{1/4}.
\eeq

Applying Laplace's method with $\Delta x=\Delta x_0$ along with a linear approximation of
$\mathcal{L}$ in the vicinity of $\Delta x\gtrsim \Delta x_0$, we obtain:
\beq
P(\Delta x>\Delta x_0;{\rm coexist})\approx \frac{1}{\tau_{\rm
    sw}} \frac{1}{\sqrt{2\pi
    16(d-1)D{t^*}^3/(3T^2)}} \frac{\sqrt{2\pi}}{\sqrt{\partial^2_t
    \mathcal{L}(t^*, \Delta x_0)}} \frac{1}{\partial_x \mathcal{L}(t^*, \Delta x_0)}
e^{-\mathcal{L}(t^*, \Delta
  x_0)}=\sqrt{\frac{3}{8}}e^{-{\left(\frac{8T^2\Delta x_0^2}{9(d-1)D\tau_{\rm sw}^3}\right)}^{1/4}}.
\eeq

Replacing $T=(r/v)R_0^{-1/M}$ and $\tau_{\rm sw}=v/(4Ds)$ yields:
\beq
P(\Delta x>\Delta x_0;{\rm coexist})\approx
\sqrt{\frac{3}{8}}\exp\left[-{\left(\frac{2^9s^3R_0^{-2/M}D^2\Delta x_0^2r^2}{9(d-1)v^5}\right)}^{1/4}\right].
\eeq

We check
self-consistently that our approximation of angular diffusion is
correct for $\Delta x_0\sim r$. The condition is met when $t^*\gg T$, or
\beq
t^*=\frac{\sqrt{3r\Delta x_0R_0^{-1/M}}}{2^{7/4}((d-1)D^2sv)^{1/4}}\gg T=\frac{r}{v}R_0^{-1/M}.
\eeq
This condition is equivalent to:
\beq
v \gg D^{2/3}r^{-1/3}.
\eeq
{\ch It} is (barely) satisfied for large population sizes
({\ch as one can check using Eq.~\ref{eq:speed} and Eq.~\ref{eq:s}}).

Finally, to get the rate of splitting events, we must multiply the
probability of a successful splitting event, $P(\Delta x>\Delta x_0;{\rm
  coexist})$, by the rate with which branches sprout from the main trunk of the phylogenic tree. Since
mutations are modeled by continuous diffusion in antigenic space,
such a new branch occurs whenever the individual virus on the trunk of
the tree (defined as the virus that will
eventually seed the entire future population) reproduces, as the two
offspring immediately become antigenically distinct because of
diffusion, and thus make two distinct branches. This happens with rate
$f(u_c)=u_c s=v^2/4D$, so that the overall rate of speciation should
scale as:
\beq
k_{\rm split}\approx
\sqrt{\frac{3}{8}}\frac{v^2}{4D}\exp\left[-{\left(\frac{2^9s^3R_0^{-2/M}D^2\Delta
        x_0^2r^2}{9(d-1)v^5}\right)}^{1/4}\right].
\eeq
Replacing $\Delta x_0=ar$, with $a$ a numerical scaling factor, gives the
result of the main text.

\section{Details of simulation implementation}

To update the fitness at each time step, we used either an exact
computation of Eq.~\ref{fit_def}, or a faster approximate method based
non-homogeneous fast Fourier transforms
\cite{keiner2009using,potts2004fast}. For the exact computation,
$c(\bx,t+\Delta t)-c(\bx,t)$ was calculated at each time step by
convolving the Kernel $H$ with $n_h(\bx,t+\Delta t)-n_h(\bx,t)$
  (exploiting the fact that this sum is sparse because not all positions get updated).
Both algorithms can be preceded by an extra-approximation for
positions $\bx'$ with $n_h(\bx',t+\Delta t)>0$ that are far enough
from the viral cloud. We approximate the Kernel $H$ by reducing
$|\bx-\bx'|$ to its projection along the direction
given by $\bx'-\< \bx\>_n$, with $\<\bx\>_n=\int d\bx\,\bx n(\bx)/N$.
This allows us to pre-compute the contributions of all these $\bx'$
to Eq.~\ref{fit_def} 
with a mere 1D convolution,
$\forall \bx$
where $n(\bx)>0$, speeding up the computation considerably.
We choose the desired combination of approximations  based on the convolution computational complexity, driven by the number of positions with $n(\bx,t+\Delta t)>0$ and $n_h(\bx,t+\Delta t)-n_h(\bx,t)>0$.
To limit errors accretion we compute the update to the convolution exactly as explained above depending on a proxy for the fitness errors.
In addition, the full convolution was recalculated with no
approximation every $10000$ steps.

\end{document}